\documentstyle[12pt]{article}
\def\be{\begin{equation}}
\def\ee{\end{equation}}
\def\ba{\begin{array}{c}}
\def\ea{\end{array}}

\begin{document}
\titlepage

   \begin{center}
{\Large \bf {New type of exact solvability
 and of a hidden nonlinear dynamical symmetry
 in anharmonic oscillators}
 }

\end{center}

\vspace{5mm}

   \begin{center}

Miloslav Znojil

\vspace{3mm}

{\small \it \'{U}stav jadern\'e fyziky AV \v{C}R, 250 68 \v{R}e\v{z},
Czech Republic\\

e-mail: znojil@ujf.cas.cz}

\vspace{5mm} and

\vspace{5mm}

Denis Yanovich

 \vspace{3mm}

{\small \it Lab. Inf. Tech., Joint Institute for Nuclear Research
(JINR) 141980,

Dubna, Moscow Region, Russia\\

 email: yan@jinr.ru  }

\end{center}


PACS {03.65.Ge}

\section*{Abstract}

{Schr\"{o}dinger bound-state problem in $D$ dimensions is considered
for a set of central polynomial potentials containing $2q$ arbitrary
coupling constants. Its polynomial (harmonic-oscillator-like,
quasi-exact, terminating) bound-state solutions of degree $N$ are
sought at an $(q+1)-$plet of exceptional couplings/energies, the
values of which comply with (the same number of) termination
conditions.  We revealed certain hidden regularities in these coupled
polynomial equations and in their roots.  A particularly impressive
simplification of their pattern occurred at the very large spatial
dimensions $D \gg 1$ where all the ``multi-spectra" of exceptional
couplings/energies proved equidistant. In this way, one generalizes
one of the key features of the elementary harmonic oscillators to
(presumably, all) non-vanishing integers~$q>0$.}

\vspace{9mm}

\noindent PACS 03.65.Ge, 03.65.Fd

 \vspace{9mm}

\newpage

\section{Introduction: quasi-exact terminating solutions}

The never-ending story of the search for exact bound-state solutions
started with the very emergence of quantum mechanics. Its part which
pays attention to the polynomial central potentials $V(r)$ in the
ordinary differential ``radial" Schr\"{o}dinger equation
 \begin{equation}
 -\psi''(r) + \frac{\ell(\ell+1)}{r^2}\,\psi(r)+
 V(r)\,\psi(r) =
 E\,\psi(r), \ \ \ \ \ \ \psi \in L_2(0,\infty)
 \label{Znojil:SE}
 \end{equation}
is not much younger.  Indeed, the elementary nineteen-century
mathematics proves sufficient for the construction of $\psi(r)$ from
(\ref{Znojil:SE}) in analytic form with, say, a power-series ansatz
for components $A(r)$ and $B(r)$ in
 \begin{equation}
 \psi(r) = r^{\ell+1}\,A(r)\,e^{B(r)}\,.
 \label{Znojil:QESp}
 \end{equation}
The most common harmonic-oscillator model $V^{(HO)}(r)= \omega^2r^2$
provides a particularly appealing illustration of such an approach
because the semiclassical exponent $B^{(HO)}(r)
=-\frac{1}{2}\omega\,r^2$ in eq. (\ref{Znojil:QESp}) describes the
correct {asymptotic} decrease of $\psi^{(HO)}(r)$ while the Taylor
series for $A^{(HO)}(r)$ degenerates to a polynomial as well.

A broad family of polynomial potentials admits a similar
specification of their ``asymptotically optimal" polynomial exponents
$B(r)$. {\it Vice versa}, for all the ``canonical" polynomial
WKB-like exponents
 \begin{equation}
 B^{(WKB)}(r) =
 \frac{1}{2}{\alpha}_0 r^2 + \frac{1}{4}{\alpha}_1 r^4 +
\ldots + \frac{1}{2q+2} {\alpha}_{q} r^{2q+2}
  \label{Znojil:Jost}\,
 \end{equation}
and for all the ``canonical" power-series choices of the ansatz
(\ref{Znojil:QESp}),
 \begin{equation}
\psi(r) = \sum_{n=0}^{\infty}\, h_n \,r^{2n+\ell+1}\,{\rm exp}\left [
-B_{WKB}(r)\right ]\,
 \label{Znojil:anainfhill}
 \end{equation}
potentials may be polynomials with $2q+1$ {\em arbitrary} couplings,
  \begin{equation}
 V(r)=V^{[q]}(r) =
 g_0\,r^2+g_1\,r^4 + \ldots + g_{2{q}}\,r^{4{q}+2}
  = [\Omega^{(q)}(r)]^2r^2+S^{(q)}(r)\,.
 \label{Znojil:geSExt}
 \end{equation}
The first, asymptotically dominating auxiliary factor
  \begin{equation}
 \ \ \ \ \ \ \
\Omega^{(q)}(r)
  = \alpha_0 +\alpha_1r^2 + \ldots + \alpha_q\,r^{2q}\,
 \label{Znojil:greeSEx}
 \end{equation}
is determined precisely by the $(q+1)-$plet of the WKB-related free
parameters while
 \begin{displaymath}
 S^{(q)}(r) = G_0r^2 +  G_1r^4 +\ldots + G_{q-1}r^{2q}
 \,
  \end{displaymath}
carries just the asymptotically less relevant information about the
full force~$V(r)$ at any~$q\geq 1$. All the relevant details may be
found in our older review of the related, so called Hill-determinant
bound-state method~\cite{Znojil:classif}.

Due to the one-to-one correspondence $g_{2q}= {{\alpha}_{q}}^2$,
$g_{2q-1} =g_{2q-1} (\alpha_{q},\alpha_{q-1})=
2\,{\alpha}_{q-1}\,{\alpha}_{q}, \ldots$ etc (or, in opposite
direction, ${\alpha}_{q}=\sqrt{g_{2q}}
>0$, ${\alpha}_{q-1}=g_{2q-1}/{(2\alpha_q)}$ etc.),
we may work with both the old and new couplings. Moreover, using the
trivial changes of variables in our differential equation
(\ref{Znojil:SE}) ($r^2= x$ etc, with all details described again
thoroughly in the above-mentioned review~\cite{Znojil:classif}), the
canonical potential (\ref{Znojil:geSExt}) generates the whole series
of its mathematical equivalents,
  \begin{equation}
 U^{[q]}(x) =
 f_0\,x^{-1}+f_1\,x +f_2\,x^2 + \ldots + f_{2{q}}\,x^{2{q}}\,
 \ \ \ \ \ \ \ f_{2q} > 0\,,
 \label{Znojil:feSExt}
 \end{equation}
  \begin{equation}
 W^{[q]}(z) =
 h_0\,z^{-3/2}+h_1\,z^{-1} + \ldots + h_{2{q}-1}\,z^{{q}-3/2}+
  h_{2{q}}\,z^{{q}-1}\,
 \ \ \ \ \ \ \ h_{2q} > 0\,,
 \label{Znojil:heSExt}
 \end{equation}
etc. Thus, the well known one-to-one mapping between harmonic
oscillator and Coulombic spectra of bound states exemplifies the
transition from (\ref{Znojil:geSExt}) to (\ref{Znojil:feSExt}) at
$q=0$. Similarly, we shall not distinguish, at any $q \geq 0$,
between the wave functions pertaining to the symmetric well
(\ref{Znojil:geSExt}) and to its descendants (\ref{Znojil:feSExt}) or
(\ref{Znojil:heSExt}).

Returning to the simplest $q=0$ models, let us emphasize that they
are extremely exceptional, possessing

\begin{itemize}

\item
{\em all} their wave functions in terminating Taylor-series form
(note that their factors $A^{(HO)}(r)$ are Laguerre {\em
polynomials});

\item
{\em all} their energies in closed form (note that the HO set forms
an {\em equidistant} family).

\end{itemize}

\noindent
 As a consequence, one should not be surprised by the existence of
numerous symmetries (and even supersymmetries \cite{Znojil:CKS}) in
the underlying Hamiltonians at $q=0$.

At $q\neq 1$, many (though not all) of these symmetries become hidden
or lost (see the monograph \cite{Znojil:Ushveridze} for wealth of
details). Still, one reveals that the exceptional polynomial
solutions exist in the form
 \begin{equation}
\psi(r) = \sum_{n=0}^{N-1}\, h_n^{(N)} \,r^{2n+\ell+1}\,{\rm
exp}\left [ -B_{WKB}(r)\right ]\,
 \label{Znojil:ana}
 \end{equation}
at all the finite integers $N \geq 1$ and $q\geq 1$ (see the review
of this point in our recent paper~\cite{Znojil:gerdt}).

\section{Schr\"{o}dinger equation at large $\ell$}

For our canonical potential (\ref{Znojil:geSExt}), the use of the
quasi-exact solution ansatz (\ref{Znojil:ana}) converts the
differential equation (\ref{Znojil:SE}) in algebraic recurrences
 \begin{equation}
 \left(
  \begin{array}{lllllll}
 B_{0} & C_0&  & & && \\
 A_1^{(1)}&B_{1} & C_1&    &&& \\
 \vdots&&\ddots&\ddots&&&\\
 A_q^{(q)}& \ldots& A_q^{(1)} &
 B_{q} & C_q&    & \\
 &A_{q+1}^{(q)}& \ldots& A_{q+1}^{(1)} &
 B_{q+1} & C_{q+1}&     \\
 & &\ddots&&&\ddots&\ddots
\\
 \end{array}
 \right )
 \left (
 \ba
 h_0^{(N)}\\
 h_1^{(N)}\\
 \vdots \\
 h_{N-1}^{(N)}\\
 0\\
 \vdots
 \ea
 \right )=0
 \label{Znojil:recurrences}
 \end{equation}
with coefficients
\begin{equation}
\begin{array}{c}
   C_n = (2n+2)\,(2n+2\ell+3),\ \ \ \ \ \
  B_n = E-\alpha_0\,(4n+2\ell +{3})
\\
 A_n^{(1)} = -\alpha_1\,(4n+2\ell+1) + \alpha_0^2 -g_0, \ \ \ \ \ \ \
 A_n^{(2)} = -\alpha_2\,(4n+2\ell-1) + 2 \alpha_0\alpha_1 -g_1,  \\
 \ldots,\\
 A_n^{(q)} = -\alpha_q\,(4n+2\ell+3-2q) +
 \left (\alpha_0\alpha_{q-1}+ \alpha_1\alpha_{q-2}+ \ldots
  +\alpha_{q-1}\alpha_0
 \right ) -g_{q-1},
 \\
\ \ \ \ \ \ \ \ \ \ \ \ \ \ \ \ \
  \ \ \ \ n = 0, 1, \ldots \ .
  \end{array}
  \label{Znojil:elem2}
   \end{equation}
They form a finite set of algebraic equations which can hardly be
solved non-numerically at the generic $q$ and $N$. In most cases,
people only pay attention to their very first ``square-matrix"
special case at $q=1$ \cite{Znojil:Ushveridze}.

In what follows, let us admit an arbitrary pair of integers $q$ and
$N$ and, for simplification, accept merely the assumption that the
spatial dimension $D$ is very large. In the other words, on the basis
of the well known formula
 \begin{equation}
 \ell=\frac{D-3}{2}, \frac{D-1}{2}, \frac{D+1}{2}, \frac{D+3}{2}, \ldots
 \end{equation}
we postulate that these numbers are {\em all} very large, $\ell \gg
1$. This is a key assumption of our forthcoming considerations,
inspired by the well known fact that for {\em any} potential $V(r)$,
the practical solution of radial Schr\"{o}dinger equations is easier
in the domain of the large angular momenta (a deeper explanation may
be found, say, in the randomly selected paper~\cite{Znojil:Bjerrum}
or in many other relevant papers with citations listed therein).

\newpage

\section{Terminating solutions at large $\ell$}

In our present very specific context of the incomplete exact
solvability, we shouldn't be misled by the observation that virtually
all the contemporary $\ell \gg 1$ calculations are based on the
perturbation expansions using the ``most natural" artificial
expansion parameter $1/\ell$. Rather, we shall follow our older paper
\cite{Znojil:Kratzer} (on the $q=2$ partial solvability) as our most
relevant guidance in what follows, having in mind the use of a
generalized expansion parameter $1/\ell^{const}$.

In its spirit, our first step will consist in a re-scaling of our
over-complete linear set (\ref{Znojil:recurrences}), $Q(E) h = 0$, in
accord with the simple rule
 \begin{equation}
 h^{(N)}_n=p_n/\mu^n\ ,
 \ \ \ \ \ \ \  \mu=\mu(D) = \left ( \frac{D}{2\alpha_{q}} \right
 )^{1/(q+1)}.
 \label{Znojil:muna}
 \end{equation}
In this way, all the elements of our non-square band-matrix
``Hamiltonian" $Q(E)$ become tremendously simplified in the leading
order in $D \gg 1$. In effect \cite{Znojil:gerdt}, we then have to
solve the much easier algebraic problem with $N$ columns and $N + q -
1$ rows,
 \begin{equation}
 \label{Znojil:trap}
 \left( \begin{array}{cccccc}
 s_1 & 1 & & & & \\
 s_2 & s_1 & 2 & &  & \\
 \vdots&  &\ddots & \ddots & & \\
 s_q & \vdots &  &s_1 & N-2& \\
 N-1& s_q & &  & s_{1} & N-1 \\
 &N-2& s_q & & \vdots& s_{1}  \\
 &&\ddots&\ddots&&\vdots\\
 &&& 2 & s_q&  s_{q-1}   \\ &&&& 1 & s_q
\end{array} \right)
 \left( \begin{array}{c}
 {p}_0\\
 {p}_1\\
\vdots \\
 {p}_{N-2}\\
 {p}_{N-1}
\end{array} \right)
=
0\
 \end{equation}
where we merely re-scaled the energy $E \ (= -g_{-1})$ and couplings
$\{g_0, \ldots,g_{q-2}\}$ in linear manner,
 \begin{equation}
 g_{k-2}=
 -\alpha_{k-1}D-\frac{\tau}{\mu^{k-1}}\,s_k,
 \ \ \ \ \ \ \
 k = 1, 2, \ldots, q\ ,
 \ \ \ \ \ \
 \tau = \left ( 2^{q+2}\,D^{q}\,\alpha_{q}
 \right )^{1/(q+1)}\,.
 \end{equation}
At this stage of development, one does not see any perceivable
progress yet. The required ``multi-spectrum" of $q$ different
``multi-eigenvalues" $s_1, \ldots, s_q$ seems obtainable only by
purely numerical means at all the larger $q$ or $N$.

\newpage

\section{A brief summary of the known non-numerical results}

Besides the well known non-numerical $q=0$ solutions of the harmonic
oscillator, also the first nontrivial $q = 1$ case need not be
discussed too thoroughly. One just solves the {\em linear algebraic}
eigenvalue problem with spectrum which proves equidistant in the
limit $D \to \infty$ (see \cite{Znojil:Kyjev}). For inspiration, let
us briefly return to this $q=1$ model in more detail: In eq.
(\ref{Znojil:trap}), the unknown quantities $s=s_1$ represent either
the energies of the sextic oscillator of eq.~(\ref{Znojil:geSExt})
or, {\it mutatis mutandis}, the charges of the spiked and shifted
harmonic oscillator \cite{Znojil:Ushveridze}, the values of which
form a finite set,
  \begin{equation}
s_1 = N-1, N-3, N-5, \ldots -N+3, -N+1\,, \ \ \ \ q = 1\,.
\label{Znojil:qjeje}
 \end{equation}
We may also very quickly recollect the next $q = 2$ case where the
solution of our problem has been found and discussed thoroughly and
with direct reference to  the quartic oscillator potential
(\ref{Znojil:feSExt}) in 1999 \cite{Znojil:Kratzer}. The first
nontrivial form of our equation (\ref{Znojil:trap}) has been solved
there in closed form for so many values of $N$ that the results could
be extrapolated to all $N = 1, 2, \ldots$. In particular, the
resulting energies were shown there to form the multiplets
 \begin{displaymath}
s_1=s_2=N-1, N-4, N-7, \ldots, -K + 2, \ \ \ \ \ \ \ N = 2K,
 \end{displaymath}
 \begin{equation}
s_1=s_2=N-1, N-4, N-7, \ldots, -K , \ \ \ \ \ \ \ N = 2K+1, \ \ \ \ \
q = 2
 \end{equation}
i.e., $s= N+2-3j, j = 1, 2, …, [(N+1)/2]$ at any wave-function degree
$N=1,2,\ldots$. In the next step of development, an "optimal"
calculation method has been discovered in our subsequent study in
2003 \cite{Znojil:gerdt}. There, we succeeded in the
re-interpretation and re-calculation of all the above $q=2$ energies
as special {\em real} roots selected out of ``hidden-symmetric"
complex triplets $ E_m = s^{1/3}e^{2\pi m/3 }, \ m = 1,2,3$. The
intermediate, auxiliary variables $s$ were produced again,
numerically, as roots of a set of polynomials $ s^6-7 s^3 -8=0$
($N=3$), $ s^10 -27 s^7 +27 s^4 -729 s=0$ ($N-4$) etc.

In the same paper, the complete solution of the next problem with $q
= 3$ has been offered. The very similar sequence of the secular
polynomials $F [s^4]$ has been obtained there, with $ F [s^4]= s^9-12
s^5 -64 s = 0$ at $N=3$, with $F [s^4]= s^16 -68 s^12 + … + 50625 =
0$ at $N=4$ etc. This gives the result
 \begin{equation}
s_2 = N-1, N-5, N-9, \ldots, N+3-4\,  \left [ \frac{N+1}{2} \right ],
\ \ \ \ \ \ \ \ q = 3, \label{Znojil:ulam}
 \end{equation}
the presentation of which proves hindered by the occurrence of the
other two independent eigenvalues $s_1$ and $s_3$. Incidentally, the
latter quantities coincide and may be specified by a closed formula.
For our present purposes it is sufficient to elucidate the
$N-$dependence of the resulting multi-spectrum via its first few
examples,
 \begin{equation}
\left | \ba s_1\\s_2\\s_3 \ea \right | =
\begin{array}{|r|r|r|r|}
-2&0&2&0\\ 2&2&2&-2\\ -2&0&2&0 \ea , \ \ \ \ \ N = 3
\end{equation}
 \begin{equation}
\left | \ba s_1\\s_2\\s_3 \ea \right | =
\begin{array}{|r|r|r|r|r|r|}
-3&-1&1&3&-1&1\\ 3&3&3&3&-1&-1\\ -3&-1&1&3&-1&1\\ \ea , \ \ \ \ \ N =
4 \end{equation}
 \begin{equation}
 \left | \ba s_1\\s_2\\s_3 \ea \right | =
 \begin{array}{|r|r|r|r|r|r|r|r|r|}
-4&-2&0&2&4&-2&0&2&0\\ 4&4&4&4&4&9&9&9&-4\\ -4&-2&0&2&4&-2&0&2&0\\
\ea , \ \ \ \ \ N = 5
 \end{equation}
etc. As long as we solved eq. (\ref{Znojil:trap}) for a sufficiently
long series of ``dimensions" $N$, we succeeded in determining the
general, extrapolated pattern for $s$. The paper itself should be
consulted for more details since the latter spectrum proves to have
an impressively compact representation in integer arithmetic, with
$s_2 = N+3-4j$, $j = 1, 2,\ldots, [(N+1)/2]$ etc.

At the time of its derivation, this feature looked ephemeral as
definitely failing to hold at the next degree $q=4$ of the potential.
At the same time, although the computer-assisted solution of eq.
(\ref{Znojil:trap}) ceased to be feasible, the $q=4$ problem looked
extremely interesting as long as it involves not only a less
appealing polynomial of the symmetric well (\ref{Znojil:geSExt}) of
the eighteen degree, but also much more interesting octic-polynomial
anharmonic oscillator (\ref{Znojil:feSExt}) and, first of all, the
phenomenologically most important case of the asymptotically cubic
force (\ref{Znojil:heSExt}).

\section{Brand new result: The case of $q=4$}

Before a thorough description of our present continuation of the
systematic and efficiently computerized symbolic-manipulation study
we should re-emphasize that at any $q \geq 2$, our algebraic set of
$N+q-1$ equations (\ref{Znojil:trap}) is nonlinear. It is formed by
the sums of the one- and two-term products of the $N+q$ unknown
quantities. In the latter role we selected the $N-1$ arbitrarily
normalized Taylor coefficients $p_j$ and the $q$ multi-eigenvalues
$s_1$, $s_2$, $\ldots s_{q}$.

In order to convey the feeling of what happens when one chooses the
different strategies of the usual elimination, we may start
experimenting at $N=1$ an find that the only real solution is
trivial, $s_1=s_2=s_3=s_4=0$. The solution at $N=2$ is also
unambiguous. Once we abbreviate $s_1=t$, $s_2=r$, $s_3=\tilde{r}$,
$s_4=\tilde{t}$, we may fix the norm by setting $p_1=1$ and proceed,
recurrently, in an upwards direction in (\ref{Znojil:trap}). This
gives $p_0=-\tilde{t}$ while $\tilde{t}^5=1$, providing finally the
unique real root $\tilde{t}=1$ and, subsequently, full solution with
$\tilde{r}=r=t=1$.

We have seen in ref. \cite{Znojil:series} that  the similar
construction is also feasible at $N=3$. Proceeding in an
upward-downward-symmetric recurent manner we now normalize $p_1=1$
and infer that $p_0=-1/t$ while $p_2=-1/\tilde{t}$. Next we
abbreviate $t\,\tilde{t} = \xi$ and re-write the remaining four lines
of eq. (\ref{Znojil:trap}) in the following form,
 \begin{equation}
 r\tilde{t}/t=\xi-2, \ \
 \tilde{r}\tilde{t}/t^2=\xi-3, \ \
 rt/\tilde{t}^2=\xi-3, \ \
 \tilde{r}t/\tilde{t}=\xi-2.
\label{Znojil:andlast}
 \end{equation}
The ratio of the two odd or two even lines eliminates $r$ or
$\tilde{r}$, respectively, and we get the same quantity
$(\xi-2)/(\xi-3)$. Its next-step elimination gives the desired
simplification $\tilde{t}^5=t^5$, with the only real solution
$\tilde{t}=t$. Then the first and last line of eq.
(\ref{Znojil:andlast}) define easily $r=r(\xi)$ and
$\tilde{r}=\tilde{r}(\xi)$ while, finally, the appropriate insertions
in one of the middle lines results in the ``secular" equation
 \begin{equation}
 {\cal P}(t) = t^3-t^2-3\,t+2=0
 \end{equation}
with the following three real roots,
 \begin{equation}
 t_1=2, \ \ \ \ \ \ t_{2,3} = \frac{1}{2}
\left ( -1 \pm \sqrt{5} \right )\,
 \end{equation}
(cf. also Table 1 below).

The elimination of the unknowns becomes almost prohibitively tedious
from $N=5$ on. The comparatively high complexity of the (necessarily,
computerized) reduction of our multi-polynomial problem
(\ref{Znojil:trap}) to the single polynomial ``secular" equation
${\cal P}(s)= 0$ is accompanied by an extremely quick growth of the
degree of our secular polynomials with $N$.  At the same time, there
exists an empirically observed fact \cite{Znojil:gerdt} that,
paradoxically, the Gr\"{o}bner-based solution of the next $q=5$
problem is in fact more easy than its $q=4$ predecessor. This
underlines the key importance of the revealed ``missing pattern" in
the $q=4$ roots as presented here in Table~1.

For compensation, the impression produced by the high degree of our
polynomials ${\cal P}(s)$ is again strongly weakened when we notice
that these functions depend in effect just on the powers of the new
auxiliary variable $z = s^{q+1}$. This has several consequences.
Firstly, we see that even if all the auxiliary roots $z$ themselves
were real, the final number of the complex roots $s$ would still be
much higher than that of their real and, hence, ``physically
acceptable" partners. Secondly, the formidable task of the search for
the real roots in the closed form did not prove to be as
prohibitively difficult as it might have appeared at first sight.

In Table~1 summarizing the results of our $q=4$ construction, a
climax of our present effort is perceived in an absolute regularity
of all its items. The pattern of extrapolation of these results
beyond their boundaries set by the computer is already fully obvious,
 \begin{eqnarray}
 s =s_4(N) = \frac{1}{2}\, \left (
P_{N} \pm \sqrt{5} \cdot Q_{}
 \right )\,,
 \nonumber\\
 \qquad {}
 P(N)=P(N)_{(j,k)}=2N+13-5j-10k\,,
 \qquad {}
  Q=Q_{(j,k)}=j-1\,,
 \nonumber\\
 \qquad {}
 \qquad {}
 \qquad {} j,k = 1, 2, \ldots\,,
 \qquad {} 2j+4k \leq N+5\,
 \label{Znojil:equation2}
 \end{eqnarray}
and does not seem to create any doubts and/or unanswered questions.
With respect to the non-doubling of the $j=1$ (i.e., $Q=0$) roots,
the elementary formula
 \begin{equation}
 total\, \# = \left (
 \begin{array}{c}
 K+1\\
 2
 \end{array}
 \right ),
 \qquad {}
 \qquad {}
 K =\left [
 \begin{array}{c}
 N+1\\
 \hline
 2
 \end{array}
 \right ],
 \end{equation}
also expresses the total number of the separate items in each column
of Table~1, i.e., of the real energy roots at each fixed $N$.

\section{Summary and outlook \label{Znojil:4}}

We reported the progress achieved in the field where the quasi-exact
solutions are sought for the radial equations where the potentials
are ``next-to-most-common". Our main result is that we were able to
construct the energies for the class of the $q=4$ models which
involves the important and very popular cubic and octic anharmonic
oscillators.

Our main task lied in the necessity of making the form of our exact
and polynomial wave functions $\psi(r)$ closed and explicit for {\em
all} their integer degrees $N=1,2, \ldots$. The main difficulty in
this direction emerges from the fact that the dimensions of the
matrices we need (or degrees ${\cal N}(N)$ of the ``effective"
secular polynomials) seem to grow extremely quickly with $N$.
Unfortunately, we did not find any regularity in the series ${\cal
N}(5)=70$, ${\cal N}(6)=126$, ${\cal N}(7)=210$, ${\cal N}(8)=330$,
${\cal N}(9)=495$, ${\cal N}(10)=715$ etc.

Our task was quite challenging formally, and we must admit that we
did not even expect that its solution could appear very soon. Our
biggest surprise occurred in the form of the explicit factorizability
of all the polynomials ${\cal P}(s)$ over the (sometimes called
``surdic") field of the quasi-complex numbers $ a+b\,\sqrt{5}$ with
rational coefficients.

One cannot resist to re-emphasize here that after a certain suitable
re-numbering and re-grouping of levels, the spectrum of our
``solvable" $q=4$ couplings/energies remains expressible directly in
terms of integers. Such a type of a generalized equidistance
re-emerges also in the next, $q = 5$ case (i.e., for the class of
potentials involving the square-root-power-series form of the quartic
oscillator, etc). The analysis of $q = 5$ already lies beyond the
scope of our present study. Even in the purely formal setting, it
lies on the very boundary of the capacity of the computers and
software which are at our disposal at present. We were still able to
factorize the corresponding effective secular polynomials at a few
$N$ in ref. \cite{Znojil:gerdt}, and we obtained the regular recipe
$s_5 = N-1, N-2, N-3, \ldots, -N+1$ there. One feels how this
achievement was formidable since at $N = 7$, the extreme coefficient
$c$ in the secular polynomial $F(s) = s^{127} - 60071 s^{121} +
\ldots + c s $ possesses as many as 72 decimal digits and, hence,
looks like a candidate for being placed in the Guiness' book of
records in the factorization context.

In conclusion, let us point out that our results sample a nice
mathematics in interplay with a useful physics. Thus, in physics, the
equidistance and representation of the energies in integer
arithmetics in the $D \to \infty$ limit will enable us to work, in
any ``realistic" dimension $D< \infty$, with perturbation theory {\em
without rounding errors}. In mathematics, the ease of the
factorization of polynomials almost certainly reflects a hidden
symmetry of the Schr\"{o}dinger equation, but in the light of the
nonlinearity of its present ``algebraization", we still do not dare
to predict any form of its possible ``explicit manifestation" in the
future.

Besides that ``new horizon", let us also stress once more that our
present study has been motivated by the disturbing paradox (revealed
in \cite{Znojil:series}) that ``phenomenologically the simplest"
cubic oscillator (such that $V(x) \approx x^3$ for $x\gg 1$) belongs,
in terms of mathematics, among ``the most difficult" examples when
its incomplete but exact $D \gg 1$ solvability is concerned. In this
sense, we described here a resolution of this paradox, showing that
the existence of the elementary and exact wave functions $\psi(x)$ in
the large$-D$ regime {\em and for any degree $N$ } is admitted not
only by the standard anharmonic Schr\"{o}dinger equation with $q=2$
(involving the quartic potentials) but also by its cubic analogue
with $q=4$.

\subsection*{Acknowledgements}

M. Z.  appreciates the support by the grant Nr. A 1048302 of GA AS
CR. The contribution of D. Y. was supported in part by the grant
01-01-00708 from the Russian Foundation for Basic Research and by the
grant 2339.2003.2 from the Russian Ministry of Industry, Science and
Technologies.



Table 1. Columns of energy-roots $s_4 =s_4(N) = \frac{1}{2}\, \left (
P_{N} \pm \sqrt{5} \cdot Q_{}
 \right )$ at $q=4$.

$$
\begin{array}{||l||r|r|r|r|r|r|r|r|r|r|r|r|r||c||}
\hline \hline
 &\multicolumn{13}{|c||}{}&\\
 &\multicolumn{13}{|c||}{P_{N}}&\  Q_{} \\
 &\multicolumn{13}{|c||}{}&\\
 \hline \hline
N=&\ 1&\ 2&3&\ 4&5&6&7&8&9&10&11&12&\ldots&all\ N\\ \hline \hline
&&&&&&&&&&&&&&\\ &0&2&4&6&8&10&12&14&16&18&20&22&\ldots&0\\
&&&&&-2&0&2&4&6&8&10&12&\ldots&0\\ &&&&&&&&&-4&-2&0&2&\ldots&0\\
&&&&&&&&&&&&&\ldots& \vdots\\ \hline &&&&&&&&&&&&&&\\
&&&-1&1&3&5&7&9&11&13&15&17&\ldots&1\\
&&&&&&&-3&-1&1&3&5&7&\ldots&1\\ &&&&&&&&&&&-5&-3&\ldots&1\\ \hline
&&&&&&&&&&&&&&\\ &&&&&-2&0&2&4&6&8&10&12&\ldots&2\\
&&&&&&&&&-4&-2&0&2&\ldots&2\\ &&&&&&&&&&&&&\ldots&\vdots\\ \hline
&&&&&&&&&&&&&&\\ &&&&&&&-3&-1&1&3&5&7&\ldots&3\\
&&&&&&&&&&&-5&-3&\ldots&3\\ \hline &&&&&&&&&&&&&&\\
&&&&&&&&&-4&-2&0&2&\ldots&4\\ &&&&&&&&&&&&&\ldots&\vdots\\ \hline
&&&&&&&&&&&&&&\\ &&&&&&&&&&&-5&-3&\ldots&5\\ \hline
&&&&&&&&&&&&&\ldots&\vdots\\ \hline \hline {\rm total}\ \#
&1&1&3&3&6&6&10&10&15&15&21&21&\ldots&\vdots\\ \hline \hline
 \ea $$


\begin{thebibliography}{99}
\footnotesize



\bibitem{Znojil:classif}
Znojil~M., Classification of oscillators in the Hessenberg-matrix
representation, {\it J.~Phys.~A: Math.~Gen.}, 1994, V.27, 4945-4968.

\bibitem{Znojil:CKS}
Cooper F., Khare A. and Sukhatme U., Supersymmetry and quantum
mechanics, {\it Phys. Rep.}, V.251, N~5,6, 267-385.


\bibitem{Znojil:Ushveridze}
Ushveridze A. G., 1994  Quasi-exactly solvable models in quantum
mechanics, Bristol, IOP Publishing, 1994.

\bibitem{Znojil:gerdt}
Znojil~M., Yanovich D. and Gerdt V. P., New exact solutions for
polynomial oscillators in large dimensions, {\it J.~Phys.~A:
Math.~Gen.}, 2003, V.36, 6531-6549.


\bibitem{Znojil:Bjerrum}
Bjerrum-Bohr N. E. J., $1/N$ expansions in nonrelativistic quantum
mechanics, {\it J. Math. Phys.}, 2000, V.41, N~5, 2515-2536.

\bibitem{Znojil:Kratzer}
Znojil M., Bound states in the Kratzer plus polynomial potentials and
the new form of perturbation theory, {\it J. Math. Chem.}, 1999,
V.26, 157-172.

\bibitem{Znojil:Kyjev}
Znojil M., Nonlinearized perturbation theories, in proceedings of The
First Int. Conf. ``Symmetries in Nonlin. Math. Physics" (3--8 July,
1995, Kyiv): {\it J. Nonlin. Math. Phys.}, 1996, V.3, 51-62.

\bibitem{Znojil:series}
Znojil~M., New series of elementary bound states in multiply
anharmonic potentials, {\it LANL preprint}, 2003, V.quant-ph,
N~arXiv: quant-ph/0304170.

\end{thebibliography}
\end{document}